\documentclass[11pt,twoside]{article}

\usepackage{asp2004}
\usepackage{epsf}

\begin{document}

\title{Diagnostics of Disks Around Hot Stars}

\author{David H. Cohen} \affil{Department of Physics \& Astronomy,
Swarthmore College, Swarthmore, PA 19081, USA}

\author{Margaret M. Hanson} \affil{Physics Department, University of
Cincinnati, Cincinnati, OH 45221, USA}

\author{Richard H. D. Townsend} \affil{Bartol Research Institute,
University of Delaware, Newark, DE 19716, USA}

\author{Karen S. Bjorkman} \affil{Department of Physics \&
Astronomy, University of Toledo, Toledo, OH 43606, USA} 

\author{Marc Gagn\'{e}} \affil{Department of Geology \& Astronomy,
West Chester University, West Chester, PA 19383, USA}

\begin{abstract}
We discuss three different observational diagnostics related to disks
around hot stars: absorption line determinations of rotational
velocities of Be stars; polarization diagnostics of circumstellar
disks; and X-ray line diagnostics of one specific magnetized hot star,
$\theta^1$ Ori C.  Some common themes that emerge from these studies
include (a) the benefits of having a specific physical model as a
framework for interpreting diagnostic data; (b) the importance of
combining several different types of observational diagnostics of the
same objects; and (c) that while there is often the need to
reinterpret traditional diagnostics in light of new theoretical
advances, there are many new and powerful diagnostics that are, or
will soon be, available for the study of disks around hot stars.

\end{abstract}

\section{Introduction}

Disks are a spatial and dynamic phenomenon, which would, ideally, be
studied by direct imaging and {\it in situ} measurements to learn about
their spatial extent and physical properties (temperature, density,
ionization structure, scale height, velocity fields).  Needless to
say, direct imaging and {\it in situ} measurements will not be
practical any time in the near future, although indirect imaging, via
interferometry, has begun to be a reality. Thus, we must use the
traditional tools of stellar astrophysics: photometry, spectroscopy,
and polarimetry in order to learn about the important physical
properties of the disks around hot stars.

While employing these diagnostics, it is important to bear in mind
several points, most if not all of which flow from our emphasis on
deriving information about the specific {\it physical} properties of
these disks:

\begin{itemize}
\item It is important to not put the cart before the horse and make
  observations simply because new instruments become available.  One
  must first determine what information is needed, preferably to test
  competing hypotheses, and then find the right diagnostic or set of
  diagnostics that can address the specific question.

\item Interpreting the results of diagnostics is more straightforward
  when there is an underlying physical model, but it is also important
  to keep in mind that the physical model might be wrong or at least
  too simplistic.  Theorists are most useful when their theories can
  make clear, quantitative predictions that can be tested by
  observations.

\item It is important to look at the sensitivity of diagnostics to
  relevant parameters and to critically evaluate how much
  diagnostic discrimination is possible from a given observation.

\end{itemize}

There are important new diagnostics coming on line in the
next few years, and our community should plan on exploiting them
effectively.  These include sensitive polarization measurements of
Zeeman splitting and new interferometers, among others.  But we should
also bear in mind that some older and more traditional diagnostics can
still be used profitably, and that the interpretation of old
measurements can also change with new insights from theorists.

In the rest of this paper, we report on three different pieces of work
involving diagnostics of hot star disks.  In \S 2, Richard Townsend
evaluates the methods for inferring stellar rotational velocities from
observed line profiles.  In \S 3, Karen Bjorkman discusses
spectropolarimetric diagnostics of hot star disks.  In \S 4, Marc
Gagne combines X-ray spectroscopy with several other diagnostics to
infer the physical properties of, and causal mechanisms at work in,
the young magnetized O star, $\theta^1$ Ori C.  We offer some
concluding remarks in \S 5.

\section{Absorption Line Diagnostics: Stellar Rotational Velocities}

Though not directly a diagnostic of hot star disks, photospheric
absorption line diagnostics of rapidly rotating hot stars have a
direct bearing on the formation mechanisms of Be disks.  If these
stars are typically more than 100 ${\rm km\,s^{-1}}$ shy of their
critical velocities, $v_{\rm crit}$, as is commonly assumed, then the
mechanism(s) required to launch material off of their surfaces and
into orbit have much more extreme energy and angular momentum
requirements compared to the situation if these stars are actually
rotating much closer to critical.  This subject is an example of how
theory and modeling can be crucial in interpreting diagnostics, even
relatively old and well-established ones.

\begin{figure}
\plotone{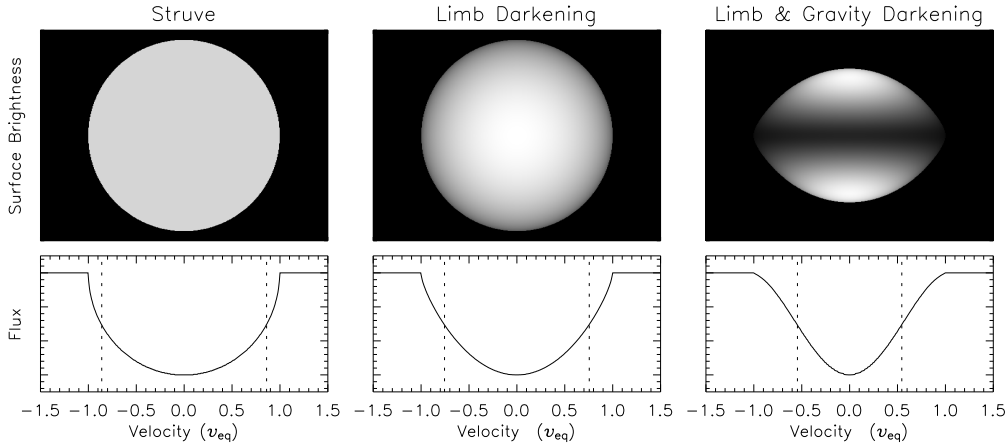}
\caption{Schematic images three stars: uniform disk (left), limb
darkened circular disk (center), oblate, gravity-darkened, and
limb-darkened disk (right).  The brightness is proportional to the
emergent flux. The lower panels are the corresponding absorption line
profile for each of the three emission models.  The separation of the
vertical lines indicate the full-width at half maximum (FWHM) of each
profile.}
\label{fig:rich_maps}
\end{figure}

The general procedure for inferring stellar rotational velocities is
to obtain stellar absorption line profiles, measure the Doppler
broadening, infer the projected equatorial rotational velocity $v_{\rm
e}\sin i$, and then apply statistical analysis to derive the
distribution of $v_{\rm e}$.  Surveys suggest Be stars have values of
$v_{\rm e} \approx 0.7$ to $0.8\,v_{\rm crit}$. However, this
procedure requires a {\it model} of the broadening process. The
original method goes back to \citet{Struve}, who assumed the intrinsic
emission profile from each surface patch is a delta function and
ignored limb darkening (we will refer to this as the ``Struve
model'').  In this case, isovelocity contours are bands on the star
parallel to the rotation axis, and the line profile mirrors the shape
of the star, as in the left-hand panels of Fig.\ \ref{fig:rich_maps}.


Observationally, the width of a given line can be parameterized using
a measure like the FWHM or the Fourier width.  In Struve's model,
these measures of the width are directly proportional to $v_{\rm
e}\sin i$, and thus measurements can be easily and unambiguously
inverted to obtain a value for $v_{\rm e}\sin i$.

We can go beyond Struve's model and include limb darkening, which
reduces the contribution from the edges of the stellar disk and thus
results in a narrower, V-shaped profile than in Struve's model, as we
show schematically in the middle panels of Fig.\ \ref{fig:rich_maps}.
Likewise, intrinsic broadening also will affect the line shape.  Even
with both limb darkening and intrinsic broadening in the model, the
correlation between line width and equatorial projected rotational
velocity is linear and thus measured values of line widths can be
inverted to derive $v_{\rm e}$.

\begin{figure}
\plotone{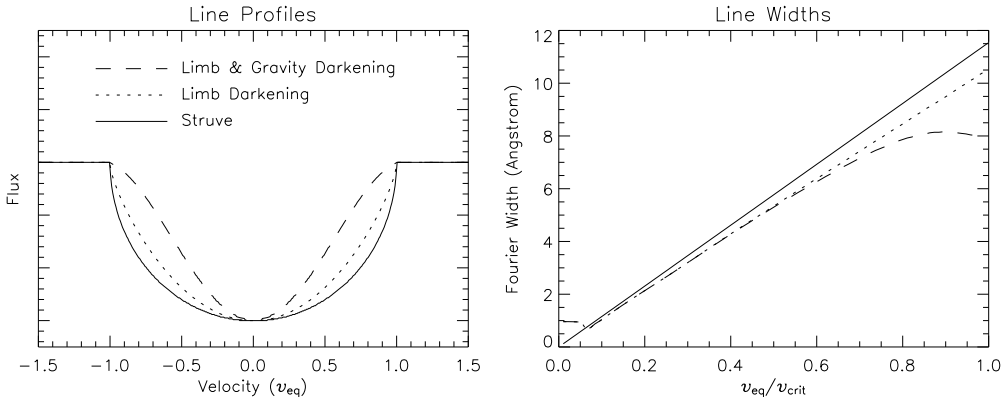}
\caption{Left: Direct comparison of the three types of line profile --
Struve, limb darkened, and gravity-darkened plus limb darkened --
shown in Fig.\ \ref{fig:rich_maps}. Note the progression to a narrower
line as each subsequent effect is included. Right: The Fourier line
width as a function of $v_{\rm eq}/v_{\rm crit}$, for the three
models. The width data for the non-Struve models are based on NLTE
line synthesis calculations for He~I 4471 \AA.  
\vspace{-0.16in}}
\label{fig:rich_profiles_width}
\end{figure}

Now, for rapidly rotating stars, one must also account for gravity
darkening.  According to von Zeipel's law, the local flux is
proportional to surface gravity.  This effect thus reduces the
contribution of the high-velocity equatorial regions to the overall
line profile.  We show this schematically in the right-hand panels of
Fig.\ \ref{fig:rich_maps}, where it can be seen that gravity darkening
further narrows the photospheric absorption line. The three different
line profile models are compared directly in the left-hand panel of
Fig.\ \ref{fig:rich_profiles_width}.


Inversion of measured line widths now becomes tricky, as the
relationship between the line width and the projected equatorial
rotational velocity is no longer linear, as we show in the right-hand
panel of Fig.\ \ref{fig:rich_profiles_width}.  Indeed, beyond $v_{\rm
eq} \approx 0.8\,v_{\rm crit}$, any increase in $v_{\rm eq}$ is
accompanied by almost no change in the line width. This strong
degeneracy means that classical techniques for $v_{\rm eq}\sin i$
determination, which neglect gravity darkening, will certainly
underestimate the true projected equatorial velocity. Since a
preponderance of surveys to date have used such classical approaches,
it appears not unreasonable that the extant data could support an
upward revision to $v_{\rm eq} \approx 0.95\,v_{\rm crit}$ for the
majority of Be stars.

In order to address whether most Be stars are rotating this rapidly,
we need multi-line diagnostics.  Different lines from different
ionization stages will sample different regions of a gravity-darkened
star.  We would expect lines with source functions weighted toward
cooler temperatures to be formed preferentially near the equator while
lines with source functions weighted toward hotter temperatures would
be formed primarily near the poles \citep{TOH2004}.  There are other
diagnostic signatures of oblateness and the associated gravity
darkening, including increased UV flux that should be seen in spectra
that cover far-UV wavelengths and photometric signatures of cooler
equators and hotter poles.  Interpreting the data, however, is
difficult, as there is a need for accurate NLTE line-blanked model
atmospheres in order to leverage these diagnostics.  It is also
possible that the observed tendency of Be stars to lie above the
photometric main sequence is a hint that extreme gravity darkening
must be taken into account when analyzing the spectral energy
distributions of Be stars \citep{ZB1991,Stoeckley1968}.

This example of the problem of inferring stellar rotational velocities
from observations of photospheric absorption line widths has direct
bearing on disk formation in Be stars, as many more mechanisms become
viable if the material feeding the disks of these stars is already
moving at $\approx 0.95\,v_{\rm crit}$.  And it also demonstrates how
even straightforward diagnostics must be tied to realistic physical
models if they are to be useful.  And once that is done, how we must
carefully test the sensitivity of observable quantities to physically
meaningful model parameters.

\section{Polarization Diagnostics for Circumstellar Disks}

\vspace{0.1in}

Polarization diagnostics are powerful because they provide relatively
direct information about the physical characteristics of circumstellar 
disks, even when these disks are spatially unresolved.  In the case of 
hot stars, where dust is generally not present, the polarization is produced
when starlight undergoes electron scattering in a non-spherically-symmetric 
circumstellar envelope (CSE).  In a sense, polarization diagnostics 
are akin to the use of stars as ``flashlights'' for illuminating the 
interstellar medium (ISM) in absorption line studies of the
ISM.  Because the polarization signal arises from the scattering
process, it represents a means for separation of the signature of the
circumstellar material from that of the star itself.  However,
polarization diagnostics for hot star disks generally are not easy to 
interpret, as the polarization percentage is often low, and some 
modeling is required to interpret the observations. Solutions are 
not always unique, and it is always necessary to correct for the interstellar 
contribution (if any) to the polarization signal.

The best approach for utilizing polarization diagnostics is to combine
polarization observations (preferably from different wavelengths) with 
other types of observational techniques, such as spectroscopy, 
photometry, interferometry, and imaging.  Different wavelength regions
probe different spatial regions, allowing for a more consistent
picture of the circumstellar environment.  Combination with other
techniques can often provide information that will distinguish between
potential non-unique solutions to polarization models.  In this section 
we briefly discuss imaging polarimetry, spectropolarimetry, and the 
removal of interstellar polarization.

Although synchrotron radiation, transport through magnetically aligned
grains in the ISM, and dust scattering all lead to polarization, in
this work we will focus primarily on electron (Thomson) scattering,
which is responsible for the polarization from hot star disks, and
which is inherently wavelength independent.  We also focus on linear
polarization only (circular polarization diagnostics of magnetic
fields are discussed elsewhere in these proceedings), using the Stokes
parameter formalism \citep{Stokes1852,Chandrasekhar1946}; for a good
discussion, see \citet{Tinbergen1996}.  The observables (as a function
of wavelength) are the polarization position angle and the net
polarization level (in percent).  The polarization level can be quite
small (1-2\% or less) if the polarized signal of scattered light is
diluted by the unpolarized direct light from the star, as is the case
for most hot star disks.

Imaging polarimetry is a useful diagnostic of YSOs with disks.  Images
of these objects are dominated by scattered light (e.g.,
\citet{Burrows1999, Wood1998}, especially when the viewing angle is
near the disk plane.  Monte Carlo modeling of linear polarization maps
can provide a tool for determining the inclination of the disk
\citep{WW2002}.  For detection purposes, as shown by
Wisniewski et al.\ (these proceedings), imaging polarimetry of
multiple point sources in a field, when combined with H-alpha emission
selection criteria, is an effective means for identifying candidate
disk systems.

\begin{figure}
\plotone{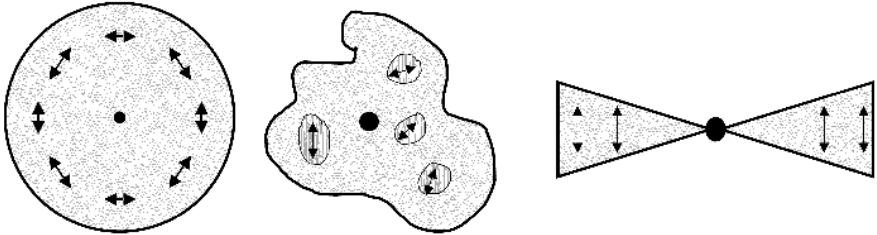}
\caption{Schematic view of polarization production in various CSE
geometries.  When the CSE is unresolved, the net polarization observed
will be the co-addition of all polarization produced.  Left: A
spherical, homogeneous distribution of scatterers will produce zero net
polarization.  Center: A non-spherical, non-homogeneous, ``blobby''
distribution of scatterers may produce a net polarization, but it will
depend on the number, spatial distribution, and relative densities of
the blobs.  If the blobs are time-variable, the observed polarization
will also be variable.  Right: A disk-like distribution of scatterers
will produce a net polarization with a position angle perpendicular to
the disk, because there is little or no cancellation due to polar
material.}
\label{fig:ksb1_cartoon}
\end{figure}

Linear polarization, caused by electron scattering in CSEs,
generates a polarization vector that is perpendicular to the
orientation of the disk (except if the disk is very optically thick).
This is shown schematically in Fig.\ \ref{fig:ksb1_cartoon}.
Since the CSEs are not spatially resolved, the observed signal is the
coadded net polarization.  As a result, the polarization position
angle gives a measurement of the orientation of the disk on the
sky, even when the disk is not resolved.

While the electron scattering polarization is wavelength independent,
the observed polarization does show a wavelength dependence.  This
happens because of the pre- and post-scattering attenuation of
polarized light by the disk material as the photons work their way
through the disk \citep{WB1995,Wood1996a,Wood1996b}.  When the
absorption cross sections are larger, less polarized flux will escape
from the disk, and so the net polarization measured will be smaller at
those wavelengths.  This effect means that the polarization
``spectrum'' can provide a direct probe of the physical conditions of
the disk material, including temperature and density.

\begin{figure}
\plotone{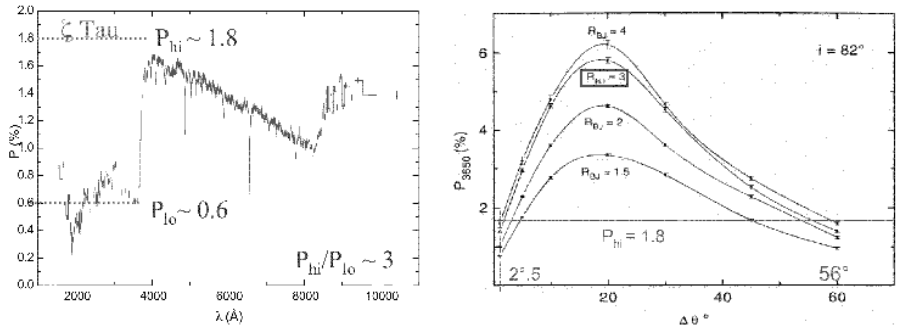}
\caption{Left: Spectropolarimetric data from the UV through the
optical for the Be star $\zeta$~Tau. Right: Monte Carlo calculations
of the peak polarization level as a function of disk opening angle for
several values of the Balmer jump polarization ratio.  Figures adapted
from \citet{WBB1997}.  The inclination angle of the disk was assumed
to be 82 degrees (nearly edge-on), consistent with the location of
$\zeta$~Tau in the ``triangle diagram'' of \citet{CW1987}.}
\label{fig:ksb2_ztau}
\end{figure}

Thus, the measured intrinsic polarization depends on the disk
geometry, density, and temperature (which affects the opacity).  As an
example, we show an observation of $\zeta$ Tau in Fig.\
\ref{fig:ksb2_ztau}, along with Monte Carlo model calculations of the
percent polarization versus disk opening angle.  As this figure shows,
there is a degeneracy in the opening angle determination.  However,
when combined with H$\alpha$ interferometry \citep{QBB1997}, the large
opening angle can be ruled out, and the small opening angle value is
determined to be the correct one.

The results from the types of studies described above show that Be
star disks are surprisingly thin, with typical opening angles of only
2.5 degrees.  These opening angles are consistent with disk
thicknesses that would be predicted by models of pressure-supported
(Keplerian) disks, although the polarization observations do not
themselves provide direct evidence that the disks are Keplerian.
Studies of the opacity of the disk material from spectropolarimetry
allow a determination of temperature and disk densities.  Derived disk
temperatures for the inner region of the circumstellar disk (where
most of the polarization is produced) are typically 75\% of the
stellar effective temperature \citep{BBW2000}.  These results show how
multi-wavelength, multi-technique observations and diagnostics
supported by modeling can provide quantitative, physical information
about circumstellar environments.

It is also important to note that models developed for the purpose of
explaining other observed phenomena generally will also carry implications 
for expected polarization, whether the models were intended for this 
purpose or not!   Polarization observations thus can be used as an
additional test for the validity of such models.  An example of this is shown
by \citep{McD2000}, looking at V/R variations in H$\alpha$ line profiles
of Be stars and their interpretation in the context of models of one-armed 
spiral density waves.  They conclude that while the models do a reasonable 
job of reproducing the general character of the spectral variations, the 
density waves would also produce specific temporal variations in the polarization 
level.  The existing polarimetric data at the time of this study did not have 
sufficient time coverage or sensitivity to detect whether these polarization
variations were observed or not.

\begin{figure}
\plotfiddle{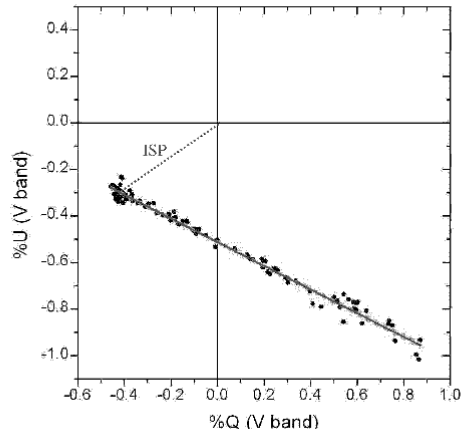}{3in}{0}{45}{45}{-144}{-10}
\caption{A Stokes QU plot showing the time variability of the V-band equivalent
polarization of the Be star $\pi$~Aqr.  When the intrinsic polarization is at its lowest
state, with the disk dissipated (or nearly so), the polarization level is only due to
the contribution of the interstellar polarization (ISP).  At this point, the ISP vector
can be determined, as indicated by the dashed line (labeled ISP) running from 
(0,0) to the ISP value.}
\label{fig:ksb3_paqr}
\end{figure}

Finally, we note that to study intrinsic polarization and fully
utilize its diagnostic power, interstellar effects need to be removed.
This is often quite difficult due to spatial variation in interstellar
polarization (ISP), distance uncertainties, and a lack of complete catalogs 
of polarization measurements for large numbers of stars all across the sky.  
One technique that can be applied to Be stars, where disks form and 
dissipate and then reform on relatively short timescales, is to monitor 
the polarization over time.  In the Q-U plane, each observation 
represents a point, and the interstellar contribution can be represented 
as a vector that shifts each point from its intrinsic value.  During monitoring, 
if observations are made at a time when the disk has dissipated
(determined based on H$\alpha$ or IR diagnostics) then the interstellar
polarization vector can be determined and (retroactively) applied to
all the other points in the Q-U plane (see Fig.\ \ref{fig:ksb3_paqr}).  
Further information about methods for removing the ISP is discussed
in some detail in \citet{QBB1997}.

Some of the prospects for the future of polarimetric diagnostics
include SPIN (SpectroPolarimetric INterferometery;
cf. \citet{SPIN2003}); polarimetric differential imaging (e.g.,
\citet{Apai2004}); Integral Field Unit (IFU) spectropolarimetry, which
allows for spectropolarimetry of many sources in one field
simultaneously; and new polarimetric capabilities on larger
telescopes.  This last item will allow for detailed polarimetry of
fainter sources and higher resolution spectropolarimetry.  For an
excellent overview of the wide range of modern polarimetric
instruments, techniques, and astronomical applications, as well as
previews of developing capabilities, the reader is referred to the
many papers contained in the proceedings of the recent conference on
Astronomical Polarimetry \citep{POLZ2004}.

\section{X-ray Line Diagnostics of the Magnetic O Star $\theta^1$ Ori C}

\vspace{0.1in}

$\theta^1$ Ori C (O7 V), the ionizing source of the Orion Nebula,
exhibits periodic H$\alpha$ emission, variable FUV P-Cygni profiles,
and X-ray emission \citep{Stahl1996,Gagne1997}. Its dipole magnetic
field strength shows the same 15.4-day periodicity, with the magnetic
pole in the line of sight at X-ray and H$\alpha$ emission maximum
\citep{Donati2002}.  These multiwavelength variability properties have
long been interpreted in terms of an oblique magnetic rotator model.

Interestingly, it wasn't until the optical spectroscopic magnetic
field measurements that the H$\alpha$ and UV observations could be
correctly interpreted, and a coherent unified picture of the
magnetically controlled circumstellar envelope of this hot star was
established. Strong, hard, variable X-ray emission was predicted by
\citet{BM1997}, assuming wind material was confined within a rigid
magnetosphere, leading to shocks and a cooling disk at the magnetic
equator. In the Babel \& Montmerle interpretation, the X-rays were
{\em weaker} when the star is viewed pole-on because of absorption in
the dense cooling disk. The later magnetic field measurements of
\citet{Donati2002} directly contradicted that prediction. It turns out
that the longitudinal magnetic field is weakest at $\phi=0.5$, when
the X-ray and H$\alpha$ emission are also at their lowest levels but
the UV absorption excess is maximized. The coherent picture that
emerges is thus one where the circumstellar material at the magnetic
equator is associated with X-ray and H$\alpha$ emission that is
occulted at $\phi=0.5$ by the star, and the wind flow along the
magnetic equator thus maximizes the UV absorption at the same phase.

\begin{figure}
\plotfiddle{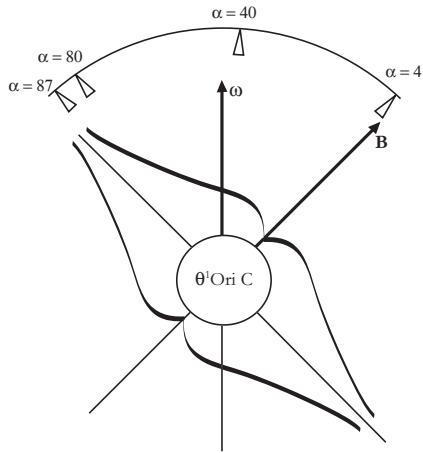}{2.3in}{0}{30}{30}{-70}{0}
\caption{Schematic of $\theta^1$ Ori C and its circumstellar
environment, with a magnetic obliquity of $\beta = 42\deg$. The
viewing angles with respect to the magnetic pole are indicated by the
four labels at the top of the diagram.  In this view, the observer
does not change position from observation to observation, but rather
the star and its magnetic field rotate, providing a fixed observer
different perspectives.  X-ray maximum occurs near viewing angle
$\alpha=3\deg$.}
\label{fig:f1}
\end{figure}

The MHD model of \citet{UO2002} was applied to this problem (see
Oksala et al. in these proceedings), showing that the radiatively
driven wind distorts the magnetic geometry, creating an outflowing
disk at large radii and strong, relatively high-density shocks above
and below the magnetic equator from $R=1.2-2{\rm R_{\star}}$. The
high-resolution spectroscopy available with the {\it Chandra} gratings
provides several useful diagnostics for analyzing the spatial extent,
temperature, and kinematics of the magnetically confined, shocked
plasma in the circumstellar environment of this star. The geometry of
the magnetosphere, as determined from the multiwavelength optical and
UV observations described above is summarized in Fig.\ \ref{fig:f1} which
also shows the viewing angle for four separate {\it Chandra}
pointings.

The most common type of X-ray spectral diagnostic applied to {\it Chandra}
grating spectra of stars is the fitting of optically thin thermal 
emission models (e.g., APEC) to the entire spectrum to determine the
column density of absorbing material overlying the X-ray emitter and
the temperature(s), abundances, and volume emission measure of the
X-ray emitting plasma. This emission measure analysis
shows very hot plasma ($T\approx30$ MK), approximately solar abundances
for most elements and substantially sub-solar Fe. The emission measure
decreases by $\approx 35$\% from maximum to minimum, but the other
parameters do not vary significantly, suggesting that the X-ray emitting 
region is close to the star ($R<1.5R_{\star}$) and is occulted by the
star as the star and the closed magnetic regions rotate once every 15.4 days.

\begin{figure}
\plotfiddle{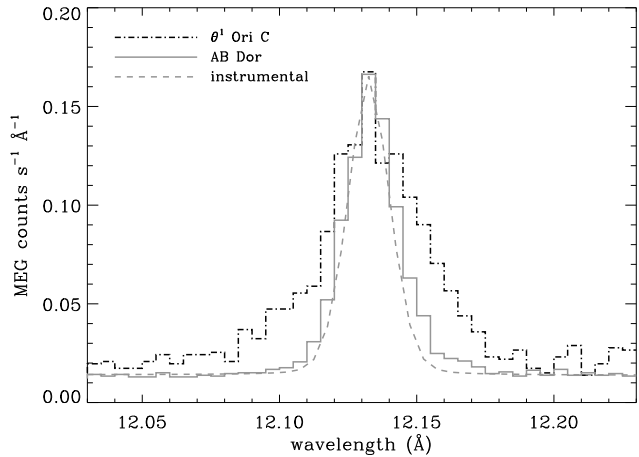}{2.3in}{0}{40}{40}{-125}{0}
\caption{The neon Lyman alpha line in the {\it Chandra} MEG spectrum of
$\theta^1$ Ori C compared to the same line seen in the MEG spectrum of the
active young M dwarf, AB Doradus.  An absolutely narrow model
convolved with the MEG instrumental response is also shown for
comparison.  The $\theta^1$ Ori C line is clearly broad.}
\label{fig:f6}
\end{figure}

X-ray emission line widths provide information about plasma
kinematics. Unlike the standard instability-driven wind shock sources
among the O stars, which have broad and asymmetric profiles,
$\theta^1$ Ori C has narrow, though well-resolved, X-ray
emission lines. As we show in Fig.\ \ref{fig:f6}, the characteristic velocity of
the emission lines is only 300 km s$^{-1}$, compared with
$v_{\infty} = 2000$~km~s$^{-1}$ seen in the stellar wind.  The very hot
plasma is moving, but at speeds much less than the ambient wind
speed. Interestingly, a few of the X-ray emission lines, from the
coolest plasma component, show larger widths, consistent with
instability-driven wind shocks.  But the harder X-rays from hotter
plasma shows behavior consistent with some degree of magnetic
confinement.

\begin{figure}
\plotfiddle{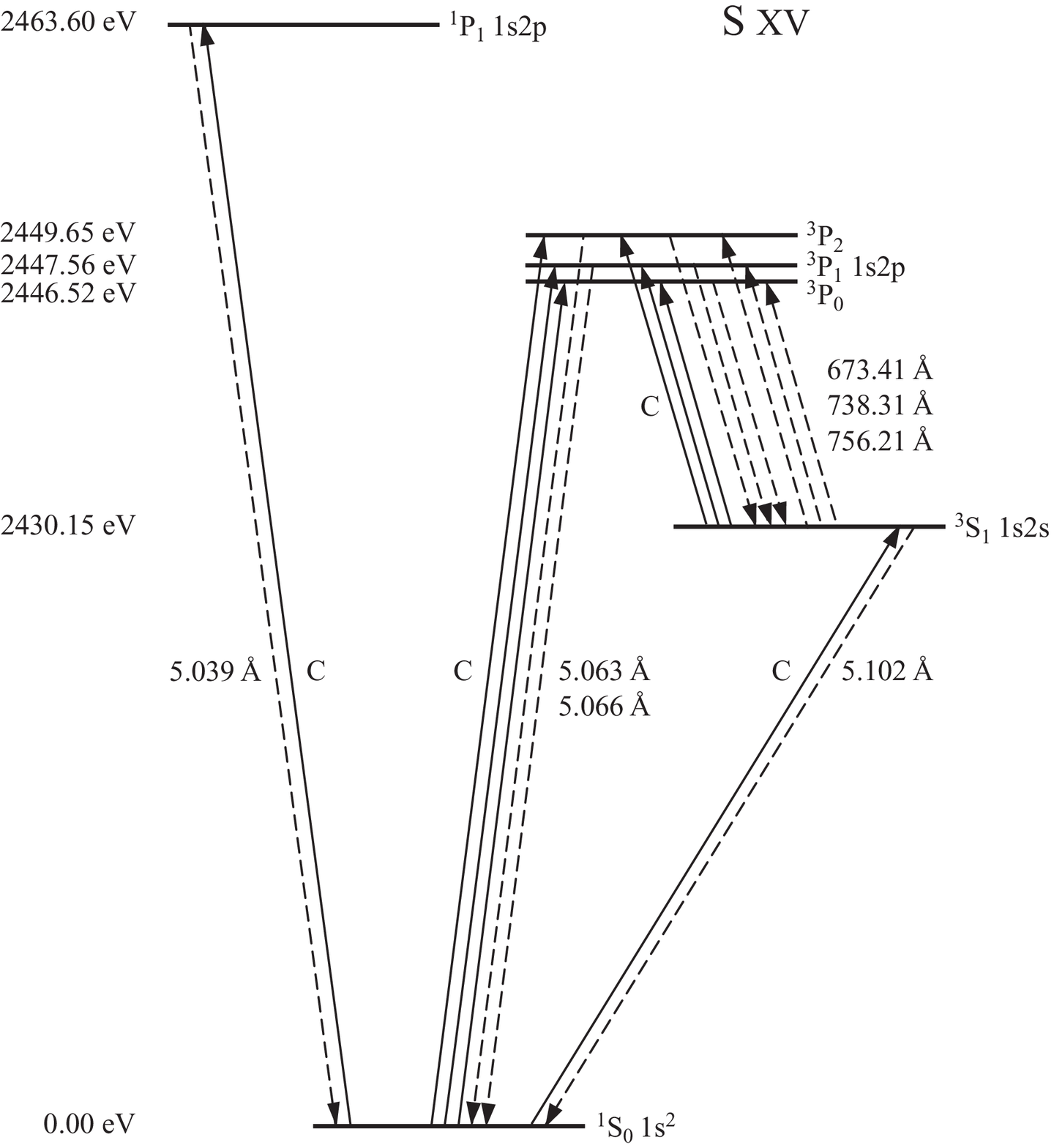}{0.8in}{0}{35}{35}{-150}{-140}
\plotfiddle{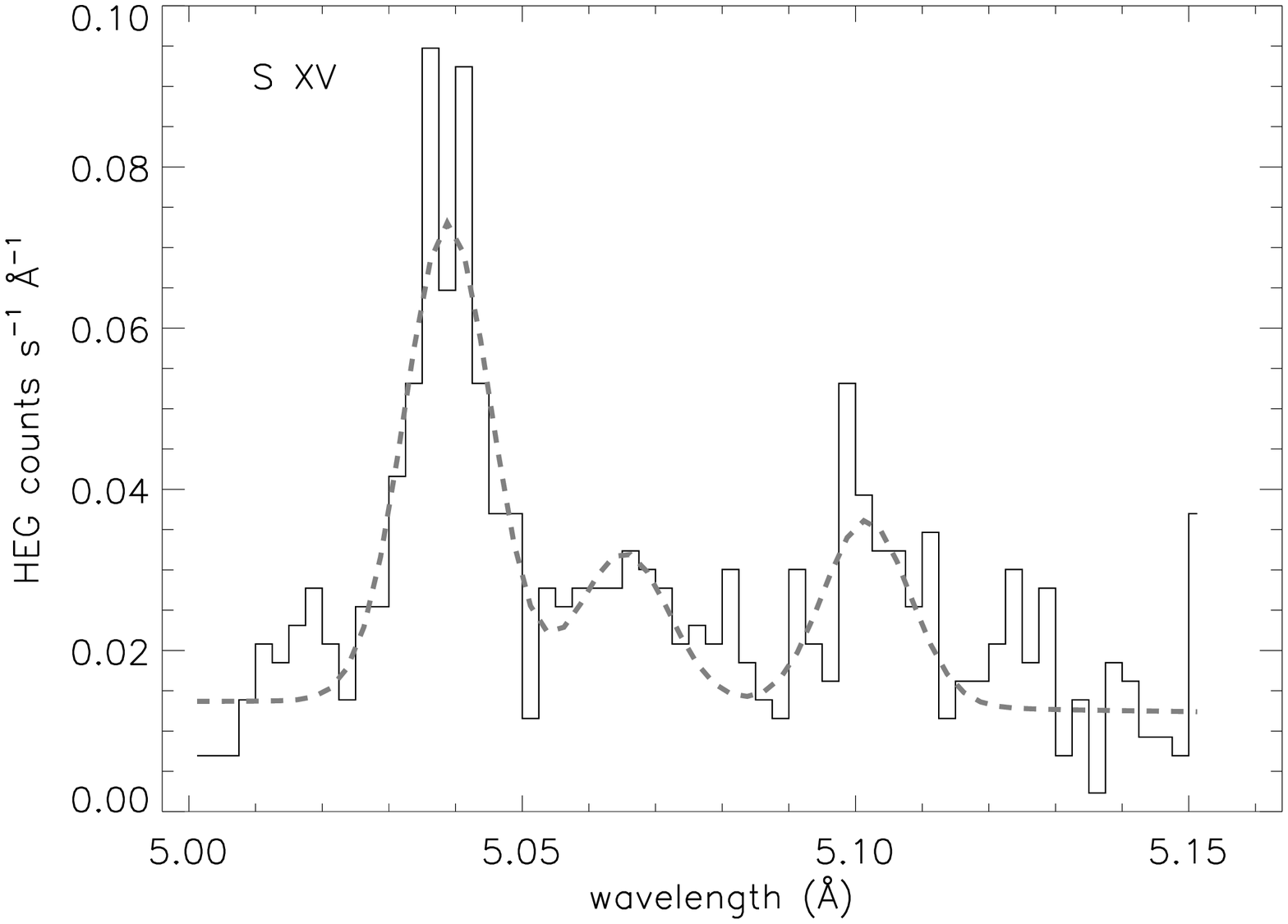}{0.8in}{0}{20}{20}{30}{30}
\plotfiddle{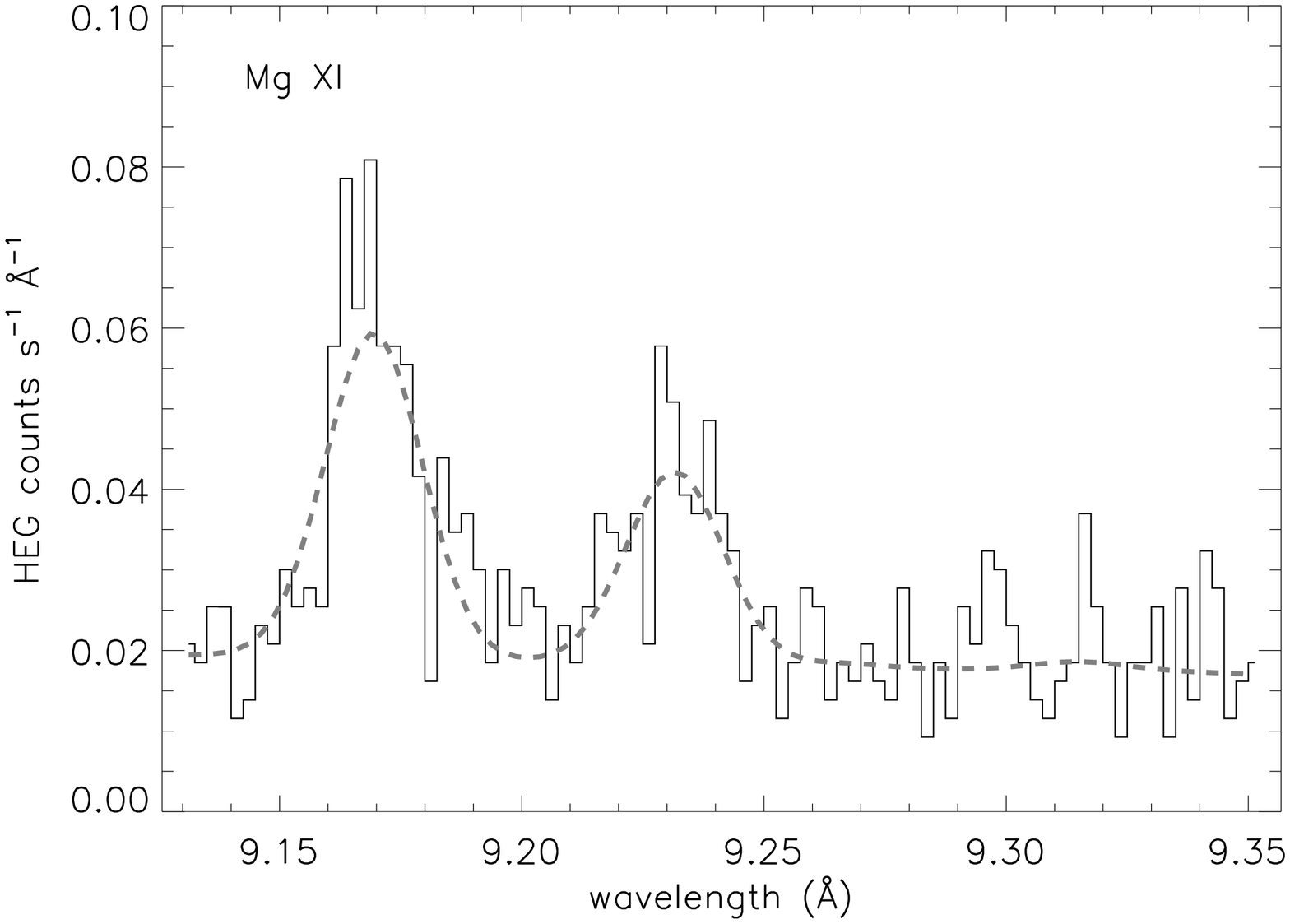}{0.8in}{0}{20}{20}{30}{0}
\caption{Energy-level diagram for S~XV (left) after \citet{GJ1969}.
Collisional excitations (C) are shown with solid lines; radiative
decays are shown with downward pointing dashed lines; photo-excitation
is shown with upward pointing dashed lines.  APED level energies and
observed transition wavelengths are also noted.  Electrons are excited
are out of the $^3S_1$ level (by collisions at high density and by
radiation close to a hot photosphere), leading to suppression of the
forbidden line (labeled by its wavelength, 5.102 \AA), and enhancing
the intercombination line (5.063, 5.066 \AA). The two panels on the
right show the observed He-like S and Mg $fir$ complexes along with
best-fit models.  The forbidden line, which is the component at the
long-wavelength end of each spectral region, is moderately strong in
the sulfur but is suppressed by photoexcitation in the magnesium.
\vspace{-0.3in}
}
\label{fig:f11}
\end{figure}

Aside from the line-profile analysis, spectral fitting, and
time-variability analysis, high-resolution X-ray spectra can provide
one other plasma diagnostic.  The ratio of
forbidden-to-intercombination ($f/i$) line strength in helium-like
ions can be used to measure the electron density of the X-ray emitting
plasma and/or its distance from a source of FUV radiation, like the
photosphere of a hot star \citep{Blumenthal1972}. The left panel of
Fig.\ \ref{fig:f11} is an energy-level diagram for S~XV similar to the
one for O~VII first suggested by \citet{GJ1969}. It shows the level
energies (eV) and transition wavelengths (\AA) of the resonance
($^1{\rm P}_1 \rightarrow ~{^1{\rm S}_0}$), intercombination ($^3{\rm
P}_{2,1} \rightarrow ~{^1{\rm S}_0}$), and forbidden ($^3{\rm S}_1
\rightarrow ~{^1{\rm S}_0}$) lines.  Aside from the usual collisional
excitations (solid lines) and radiative decays (dashed lines), the
metastable $^3{\rm S_1}$ state can be depopulated by collisional
and/or photo-excitation to the $^3{\rm P}$ states. In late-type stars,
the $f/i$ ratios of low-Z ions like N~VI and O~VII are sensitive to
changes in electron density near or above certain critical densities
(typically in the $10^{12}$~cm$^{-3}$ regime). Higher-Z ions have
higher critical electron densities typically not seen in normal
(non-degenerate) stars.  \citet{Kahn2001} have used the $f/i$ ratios
of N VI, O~VII, and Ne~IX and the radial dependence of the incident UV
flux to determine that most of the emergent X-rays from the O4
supergiant $\zeta$~Pup were formed in the far wind, consistent with
radiately-driven wind shocks.

In $\theta^1$~Ori C, the low-Z forbidden lines are completely
suppressed.  The right-hand panel of Fig.\ \ref{fig:f11} shows the
resonance, intercombination, and forbidden lines of Mg XI and Ar
XVII. Modeling the excitation kinematics of these lines at the
dominant plasma temperature of 33~MK, the undetected Mg~XI forbidden
line suggests a formation radius $R < 1.8 {\rm R_{\star}}$ while the
S~XV $f/i$ ratio provides $1\sigma$ lower and upper bounds of 1.2 and
$1.5 {\rm R_{\star}}$, consistent with Mg~XI. The Si~XIII lines have
the highest S/N, but the forbidden line is blended with a strong line
of Mg~XII. That said, the Si XIII lines and the lower-S/N Ar~XVII
lines are consistent with the S~XV measurement, i.e., less than
approximately half a stellar radius above the photosphere. This is
also consistent with the reduction in emission measure seen near phase
0.5, based on the geometry of the occultation and the depth of the dip
in the X-ray light curve.

All of these X-ray diagnostics, and the earlier optical and UV
observations, can be understood in terms of a model in which a
radiation-driven wind is channeled by a dipole magnetic field toward
the magnetic equator where flows from the two hemispheres collide and
shock-heat. This picture has been confirmed by numerical MHD
simulations. These simulations, based on those of \citet{UO2002},
show that the magnetically channeled flows are confined within about
a stellar radius of the photosphere, i.e., $R \approx 1-2 R_{\star}$,
where they shock heat plasma to approximately 30~MK, producing
relatively narrow, symmetric, X-ray emission lines.

This analysis demonstrates the power of quantitative X-ray spectral
diagnostics to determine the values of physically meaningful
parameters related to circumstellar matter in magnetic hot stars.
This situation is only possible, however, because of the existence of
quality datasets in the optical and UV to augment the X-ray data, and,
importantly, because of the numerical models of the magnetic wind of
$\theta^1$ Ori C that makes quantitative predictions that the data can
be used to test.

Finally, we note that the circumstellar envelope of this star is not a
true disk, in the sense of Be stars, for example, where the
circumstellar envelope is thin and orbitally supported. In $\theta^1$
Ori C, the envelope is more toroidal and is supported, but only
temporarily, by the magnetic field. Prospects for finding other stars
in this category - young hot stars with magnetically controlled
circumstellar matter - involve identifying hard and variable X-ray
sources, and then combining those data with H$\alpha$ emission and UV
absorption measurements. So far, about a dozen such O-, B-, and A-type
candidates, mostly spectroscopic binaries, have been identified in
the Orion Nebula region.

\section{Conclusions}

The field of diagnostics is an extraordinarily broad and diverse one,
and in this paper we have touched on only three rather specific
diagnostics and their associated applications.  However, while there
is both a certain amount of common ground as well as some interesting
differences among these three examples, several underlying lessons
emerge. First of all, researchers must be aware of their biases and
always be leery of the assumptions that underly the models that they
rely on to interpret data.  Employing models to interpret data is of
crucial importance, though, as determining physical parameters of
interest is the ultimate motivation for employing diagnostics in the
first place.

The section on absorption line profiles and Be star rotational
velocities shows that even traditional diagnostics must be tied to
models in order to provide meaningful physical interpretations, and
that the assumptions behind models should be reevaluated periodically
as theories become more sophisticated.  Also, observers must not lose
sight of which specific properties of the observational data are
sensitive to the physical parameters of interest.  The section on
polarization diagnostics demonstrates how the electromagnetic
radiation we measure contains a huge amount of information that can be
extracted if the diagnostics are interpreted in light of physical
models.  And, like so many diagnostic techniques, there are flies in
the ointment (in this case, the removal of interstellar polarization)
that can be difficult to deal with, but which must be accounted for if
meaningful conclusions are going to be drawn from the data.  This case
is especially satisfying because it is in fact one of the inherently
interesting properties of the class of objects under study -- namely
the rapid dispersal and then formation of Be disks -- that enables us
to make an accurate assessment of ISP contamination. Finally, the
section on $\theta^1$ Ori C presents a nearly unique case that
nonetheless illustrates some of the basic guidelines for employing
observational diagnostics.  In this final section, we see again how
important it is to have an underlying physical model for the
interpretation of diagnostics.  We also see how combining data at
several wavelengths is vital.  And finally we see how important
quantitative measurements of specific physical properties (e.g.\ X-ray
line widths and line ratios) are for making meaningful progress in
understanding astrophysical systems of interest.


\begin{thebibliography}{}

\bibitem[Adamson et al.(2004)]{POLZ2004} Adamson, A., Aspin, C.,
Davis, C., \& Fujiyoshi, T.  (editors) 2004, ``Astronomical
Polarimetry: Current Status and Future Directions'', ASP Conf. Proc.,
in press.

\bibitem[Apai et al.(2004)]{Apai2004} Apai, D., Pascucci, I.,
Brandner, W., Henning, T., Lenzen, R., Potter, D.~E., Lagrange, A.-M.,
\& Rousset, G.\ 2004, \aap, 415, 671

\bibitem[Babel \& Montmerle(1997)]{BM1997} Babel, J., \& Montmerle, T. 1997, \apj, 485, L29 

\bibitem[Bjorkman, Bjorkman, \& Wood(2000)]{BBW2000} Bjorkman, K.~S.,
Bjorkman, J.~E., \& Wood, K.\ 2000, IAU Colloq.~175: The Be Phenomenon
in Early-Type Stars, (ed.\ M.\ Smith and H.\ Henrichs), ASP
Conf. Ser., 214, 603

\bibitem[Blumenthal, Drake, \& Tucker(1972)]{Blumenthal1972}
Blumenthal, G.~R., Drake, G.~W.~F., \& Tucker, W.~H.\ 1972, \apj, 172,
205

\bibitem[Burrows et al.(1999)]{Burrows1999} Burrows, S. 1999, \apj,
516, L95

\bibitem[Chandrasekhar(1946)]{Chandrasekhar1946} Chandrasekhar,
S. 1946, \apj, 104, 110

\bibitem[Chesneau, et al.(2003)]{SPIN2003} Chesneau, O., Wolf, S., \&
Domiciano de Souza, A.\ 2003, \aap, 410, 375

\bibitem[Cot\'e \& Waters(1987)]{CW1987} Cot\'e, J., \& Waters,
L.B.F.M.  1987, \aap, 176, 93

\bibitem[Donati et al.(2002)]{Donati2002} Donati, J.-F., Babel, J.,
Harries, T.~J., Howarth, I.~D., Petit, P., \& Semel, M.\ 2002, \mnras,
333, 55

\bibitem[Gabriel \& Jordan(1969)]{GJ1969} Gabriel, A.~H.~\& Jordan,
C.\ 1969, \mnras, 145, 241

\bibitem[Gagn\'e et al.(1997)]{Gagne1997} Gagne, M., Caillault, J.,
Stauffer, J.~R., \& Linsky, J.~L.\ 1997, \apjl, 478, L87

\bibitem[Kahn et al.(2001)]{Kahn2001} Kahn, S.~M., Leutenegger, M.~A.,
Cottam, J., Rauw, G., Vreux, J.-M., den Boggende, A.~J.~F., Mewe, R.,
\& G{\" u}del, M.\ 2001, \aap, 365, L312

\bibitem[McDavid et al.(2000)]{McD2000} McDavid, D.A., Bjorkman, K.S.,
Bjorkman, J.E., \& Okazaki, A.T.  2000, in IAU Colloq. 175: The Be
Phenomenon in Early-Type Stars, (ed.\ M.\ Smith and H.\ Henrichs), ASP
Conf. Ser., 214, 460

\bibitem[Quirrenbach et al.(1997)]{QBB1997} Quirrenbach, A., et al.
1997, \apj, 479, 477

\bibitem[Tinbergen (1996)]{Tinbergen1996} Tinbergen, J. 1996, {\it
Astronomical Polarimetry}, (Cambridge: Cambridge Univ. Press), Ch.\ 2

\bibitem[Townsend, Owocki, \& Howarth(2004)]{TOH2004} Townsend,
R. H. D., Owocki, S. P., \& Howarth, I. D.  2004, \mnras, 350, 189

\bibitem[Shajn \& Struve(1929)]{Struve} Shajn, G., \& Struve, O. 1929,
\mnras, 89, 222

\bibitem[Stahl et al.(1996)]{Stahl1996} Stahl, O.~et al.\ 1996, \aap,
312, 539

\bibitem[Stoeckley(1968)]{Stoeckley1968} Stoeckley, T. R. 1968,
\mnras, 140, 141

\bibitem[Stokes(1852)]{Stokes1852} Stokes, G. G. 1852, {\it
Trans. Camb. Phil. Soc.}, vol. 9, pt.\ III, pp.\ 399-416.

\bibitem[ud-Doula \& Owocki(2002)]{UO2002} ud-Doula, A.~\& Owocki,
S.~P.\ 2002, \apj, 576, 413

\bibitem[Whitney \& Wolff (2002)]{WW2002} Whitney, B.A. \& Wolff, M.J,
2002, \apj, 574, 205

\bibitem[Wood \& Bjorkman(1995)]{WB1995} Wood, K., \& Bjorkman, J.E.
1995, \apj, 443, 348

\bibitem[Wood et al.(1996a)]{Wood1996a} Wood, K., Bjorkman, J.E.,
Whitney, B.A., \& Code, A.D.  1996, \apj, 461, 828

\bibitem[Wood et al.(1996b)]{Wood1996b} Wood, K., Bjorkman, J.E.,
Whitney, B.A., \& Code, A.D.  1996, \apj, 461, 847

\bibitem[Wood, Bjorkman, \& Bjorkman(1997)]{WBB1997} Wood, K.,
Bjorkman, K.S., \& Bjorkman, J.E.  1997, \apj, 477, 926

\bibitem[Wood et al.(1998)]{Wood1998} Wood, K., Kenyon, S.J., Whitney,
B., \& Turnbull, M.  1998, \apj, 497, 404

\bibitem[Zorec \& Briot(1991)]{ZB1991} Zorec, J. \& Briot, D. 1991,
\aap, 245, 150

\end{thebibliography}
\end{document}